\documentclass[a4paper,UKenglish,cleveref,numberwithinsect]{lipics-v2021}

\title{Query Languages for Machine-Learning Models}

\titlerunning{Query Languages for ML}

\author{Martin Grohe}{RWTH Aachen University, Germany}{grohe@informatik.rwth-aachen.de}{https://orcid.org/0000-0002-0292-9142}{}

\authorrunning{M.~Grohe}

\Copyright{Martin Grohe}



\funding{Funded by the European Union (ERC, SymSim,
101054974). Views and opinions expressed are however those of the
author(s) only and do not necessarily reflect those of the European
Union or the European Research Council. Neither the European Union
nor the granting authority can be held responsible for them.}

\acknowledgements{Thanks to Nicole Schweikardt, Erich Grädel, and Jan Van den Bussche for
valuable comments on an earlier version of this paper.
}

\nolinenumbers

\hideLIPIcs

\usepackage{latexsym}
\usepackage{amsmath}
\usepackage{amssymb}
\usepackage{mathtools}
\usepackage{stmaryrd}
\usepackage{BOONDOX-cal}

\numberwithin{equation}{section}

\newcommand{\CA}{{\mathcal A}}
\newcommand{\CB}{{\mathcal B}}

\newcommand{\CG}{{\mathcal G}}

\newcommand{\CN}{{\mathcal N}}

\newcommand{\CQ}{{\mathcal Q}}
\newcommand{\CR}{{\mathcal R}}

\newcommand{\CRlin}{{\mathcal R}_{\textup{lin}}}

\newcommand{\BN}{\mathbf N}
\newcommand{\BR}{\mathbf R}

\newcommand{\bigmid}{\mathrel{\big|}}
\newcommand{\Bigmid}{\mathrel{\Big|}}

\renewcommand{\tilde}{\widetilde}
\renewcommand{\hat}{\widehat}

\renewcommand{\vec}[1]{\boldsymbol{#1}}

\newcommand{\uend}{\hfill$\lrcorner$}

\DeclareMathOperator{\ar}{ar}

\newcommand{\Fraisse}{Fra\"{\i}ss{\'e}}

\renewcommand{\phi}{\varphi}
\renewcommand{\epsilon}{\varepsilon}

\newcommand{\Nat}{{\mathbb N}}
\newcommand{\PNat}{{\mathbb N}_{>0}}
\newcommand{\Real}{{\mathbb R}}

\newcommand{\Rat}{{\mathbb Q}}
\newcommand{\Rbot}{\Real_\bot}
\newcommand{\Qbot}{\Rat_\bot}

\newcommand{\logic}[1]{\textsf{\upshape #1}}
\newcommand{\FO}{\logic{FO}}
\newcommand{\FOSUM}{\logic{FO(SUM)}}
\newcommand{\IFPSUM}{\logic{IFP(SUM)}}
\newcommand{\sIFPSUM}{\logic{sIFP(SUM)}}
\newcommand{\AC}{\logic{AC}}
\newcommand{\TC}{\logic{TC}}
\newcommand{\NP}{\logic{NP}}
\newcommand{\PTIME}{\logic{P}}
\newcommand{\ifp}{\logic{ifp}}

\newcommand{\sem}[1]{\left\llbracket#1\right\rrbracket}

\DeclareMathOperator{\wt}{\mathsf{wt}}
\DeclareMathOperator{\bias}{\mathsf{bias}}
\DeclareMathOperator{\inp}{\mathsf{inp}}
\DeclareMathOperator{\relu}{ReLU}
\DeclareMathOperator{\id}{id}
\newcommand{\lein}{\le_{\mathsf{in}}}
\newcommand{\leout}{\le_{\mathsf{out}}}
 
\newcommand{\eval}{\logic{eval}} 
\newcommand{\edge}{\logic{edge}} 
\newcommand{\Eval}{\logic{Eval}} 
\newcommand{\Integrate}{\logic{Integrate}} 
\newcommand{\Zero}{\logic{Zero}} 
\newcommand{\NonZero}{\logic{NonZero}} 
\newcommand{\Irr}{\logic{Irr}} 
\newcommand{\avg}{\logic{avg}}

\newcommand{\ite}[3]{\operatorname{\mathsf{if}}#1\operatorname{\mathsf{then}}#2\operatorname{\mathsf{else}}#3}

\begin{document}

\maketitle

\begin{abstract}
In this paper, I discuss two logics for weighted finite structures: first-order logic with summation (FO(SUM)) and its recursive extension IFP(SUM). These logics originate from foundational work by Grädel, Gurevich, and Meer in the 1990s. In recent joint work with Standke, Steegmans, and Van den Bussche, we have investigated these logics as query languages for machine learning models, specifically neural networks, which are naturally represented as weighted graphs. I present illustrative examples of queries to neural networks that can be expressed in these logics and discuss fundamental results on their expressiveness and computational complexity.
 \end{abstract}

\section{Introduction}

\subsection*{Background}
In the 1980s, finite model theory \cite{EbbinghausF99,Libkin04}
emerged as a new subfield of mathematical logic motivated by the idea
of applying tools from mathematical logic to the analysis of algorithms for
and the complexity of computational problems. \emph{Finite} model theory was needed because in many cases problem instances are
 finite structures such as graphs, Boolean formulas, or circuits,
or, generically, finite strings over a finite alphabet. In principle,
the input to any computational problem can be described as a finite
string over the binary alphabet. Yet the mathematical models naturally describing computational problems often have a
numerical component that goes beyond discrete finite structures.

An important showcase for finite model theory is the theory of
relational databases \cite{AbiteboulHV95}. As a first approximation,
relational databases are finite relational structures, and (the core
of) the query language SQL is the relational calculus, that is,
first-order logic. Valuable insights, for example, about the
expressivity of query languages can be obtained from this perspective.
Yet real databases have datatypes with infinite domains, such as
integers, and query languages like SQL can express arithmetical or
other operations over these infinite domains. Constraint databases
\cite{KuperLP00} even consider finitely defined relations in an infinite
model, such as the ordered field of real numbers, and are analysed by
combining methods from finite and infinite model theory. Hybrid
dynamical systems \cite{DoyenFPP18} combine continuous dynamics with a
discrete control. They are typically described by models such as timed
or hybrid automata and analysed using logics with a discrete and
continuous component. Weighted graphs, which appear as inputs to many
optimisation problems, provide another typical example of \emph{hybrid
  models} with a discrete and a numerical component. Perhaps the most
important hybrid models in current computer-science research are
machine-learning models, specifically neural networks, which can also
be viewed as weighted graphs.

Despite the infinite, often continuous, parts that all these hybrid
models involve, the methods of finite model theory, such as logics
describing computation and capturing complexity classes, or
Ehrenfeucht-\Fraisse\ games and locality techniques for analysing
expressiveness, seem well-suited for analysing computational aspects
of the hybrid models as well. Recognising this, Grädel and Gurevich
\cite{GradelG98} systematically extended the framework of finite model
theory to such ``meta-finite models'', as they called
them. They extended
important logics of finite model theory to the meta-finite realm and
developed a basic descriptive complexity theory. Grädel and Meer~\cite{GradelM95}
further developed this descriptive complexity theory for hybrid
models over the reals, which are precisely the structures we are
most interested in here. Over the years, there
have been various other suggestions of logics for hybrid models, in
different degrees of generality and with different applications in
mind, for example,
\cite{Geerts23,GeertsMRVV21,KuskeS17,Torunczyk20,BergeremS21}.\footnote{\ldots
  just to name a few. I do not aim for completeness, and this paper is
  not intended as a survey. View it as a ``personal perspective''.
  Nevertheless, the area would deserve a more
  comprehensive treatment.}  Let me also mention the very interesting
work on logics with semiring semantics, for example,
\cite{BrinkeGMN24,DannertGNT21,GradelT25,GreenKT07}, differing from the
logics considered here in that weights and numerical values are not
addressed explicitly in the language, but only appear as semantic
values of formulas.

\subsection*{Querying ML Models}
Machine learning models are typical hybrid models in the sense just discussed. Due to their ubiquitous presence, they are being studied from many different perspectives, and I would like to argue that the perspectives of finite (or meta-finite) model theory and database theory, in particular, the idea of querying machine learning models, have the potential to offer new insights and methods.

Modern data science projects generate vast numbers of model artefacts
throughout their lifecycle, encompassing training runs, hyperparameter
configurations, architectural variations, and performance
metrics. While contemporary platforms like MLflow provide valuable
metadata-based filtering and search capabilities for organising
experiments, they offer little support for querying the internal
behaviour or semantics of the models themselves. Here ``querying'' a
model goes far beyond evaluating it on a specific input; it allows for an
interaction with the inner workings of a model. The models themselves
are complex structures. Specific queries may enable us to understand
the functionality of individual components in a model. Or, on the
semantical side, we may target the global behaviour of models,
that is, the behaviour on all inputs and not just a specific one, for example, to understand the
robustness of a model or to verify its correctness.

Crucially, we do not aim for a fixed, prescribed set of queries, but
for a flexible language that allows us to formulate queries tailored
towards specific properties of models, possibly in a specific
application context, and to adaptively refine queries to interact with
a model. This may allow for a much more refined understanding, or explanation, of machine learning models than current methods based on
fixed, generic explanation modes, for example, counterfactual
explanations of specific outcomes, offer. Querying a model is also
the basis of verification (just phrased in a different language:
querying is the database version of model checking).

In \cite{ArenasBBPS21,ArenasBBCS24}, the authors study Boolean
languages that are well-suited for querying discrete ML models such as
decision trees, and they analyse the expressiveness and complexity of
these languages. However, most ML models are hybrid models (in the
sense discussed above) with a discrete and a continuous component, and to
fully access these models, the query language should reflect this. In
Sections~\ref{sec:fosum} and \ref{sec:ifpsum}, we will see two
languages with this ability, $\FOSUM$ and $\IFPSUM$. The logics go back to the work of Grädel and Meer~\cite{GradelM95} and Grädel and Gurevich~\cite{GradelG98}; the exact versions that we consider here have recently
been introduced in \cite{GroheSSV26a,GroheSSV25}. Both of these languages
stand in the tradition of database query languages
\cite{AbiteboulHV95,ChandraH82} and finite-model theory, and they can
be seen as instantiations of Grädel and Gurevich's framework of
meta-finite model theory \cite{GradelG98}.

Let me remark that this paper's perspective on logics as query languages for ML models is
fundamentally different from the work on logical characterisations of the expressiveness of ML models such as graph neural networks and
transformers \cite{BarceloKM0RS20,Grohe21,Grohe24b,StroblMW0A24}.

\section{Preliminaries}
\label{sec:prel}

We denote the sets of real, rational, and natural numbers (including
$0$) by $\Real,\Rat,\Nat$ and let $\Rbot\coloneqq\Real\cup\{\bot\}$
and $\Qbot\coloneqq\Rat\cup\{\bot\}$, where $\bot$ is an additional
element representing undefined values. We extend the usual order $\le$
on $\Real$ and $\Rat$ to $\Rbot$ and $\Qbot$ by letting $\bot< x$
for all $x\in\Real$. We extend addition, subtraction, and
multiplication by letting $x+y\coloneqq\bot$, $x-y\coloneqq\bot$, and
$x\cdot y\coloneqq \bot$ if $x=\bot$ or $y=\bot$. Similarly, we extend
division by letting $x/y\coloneqq\bot$ if $x=\bot$ or $y=\bot$ or
$y=0$.

We use bold-face letters to denote tuples, such as $\vec
x=(x_1,\ldots,x_k)$, and we denote the length $k$ of this tuple by
$|\vec x|$.

\subsection{Weighted Structures}
\label{sec:weighted-structures}
The hybrid models that we consider here take the form of
\emph{weighted (finite, relational) structures}.  A \emph{weighted
  vocabulary} is a finite set $\Upsilon$ of \emph{relation symbols}
and \emph{weight-function symbols}. Each symbol $S$ has an
\emph{arity} $\ar(S)\in\Nat$. Note that we do admit relations and
weight-functions of arity $0$. While $0$-ary relations are not
particularly interesting, $0$-ary weight functions can be useful,
because we can view them as \emph{weight constants}.
An $\Upsilon$-structure $\CA$ consists of a finite nonempty set $A$, the
\emph{universe} of $\CA$, for each relation symbol
$R\in\Upsilon$ a relation $R^{\CA}\subseteq A^{\ar(R)}$, and for each
weight-function symbol $F\in\Upsilon$ a function
$F^\CA:A^{\ar(F)}\to\Real_\bot$. We always denote structures by calligraphic letters $\CA,\CB,\ldots$ and their universes by the corresponding roman letters $A,B,\ldots$. 

\begin{example}\label{exa:weighted-graph1}
  A weighted graph can described as a $\{\wt\}$-structure $\CG$ for a
  binary weight-function symbol $\wt$. The universe of $\CG$ is the
  vertex set of the graph, and for all vertices $v,w$, if there is an
  edge from $v$ to $w$ then $\wt^\CG(v,w)\in\Real$ is the weight of this edge,
  and otherwise $\wt^\CG(v,w)=\bot$. For a weighted graph $\CG$, we
  usually denote the vertex set (the universe) by $V^\CG$ instead of
  $G$, and we let
  $E^\CG\coloneqq\big\{(v,w)\in (V^\CG)^2\bigmid
  \wt^\CG(v,w)\neq\bot\big\}$.

  The weighted graph $\CG$ is undirected if $\wt^\CG$ is symmetric,
  and it is loop-free if $\wt^\CG(v,v)=\bot$ for all vertices $v$.
  \uend
\end{example}

For vocabularies $\Upsilon\subseteq\Upsilon'$, an
\emph{$\Upsilon'$-expansion} of an $\Upsilon$-structure $\CA$ is an
$\Upsilon'$-structure $\CA'$ with $A'=A$ and $S^{\CA'}=S^\CA$ for all
$S\in\Upsilon$. If $\CA'$ is an expansion of $\CA$, then $\CA$ is a
\emph{reduct} of $\CA'$.

Observe that standard finite relational structures are weighted
structures whose vocabulary contains no weight-function symbols. We
can think of weighted structures as 2-sorted structures consisting of
a finite relational structure as the first sort and the extended
real numbers as the second sort. We typically assume additional
structure on the reals, such as a linear order and arithmetic operations. Weight functions map elements from the first to the
second sort.

A weighted structure $\CA$ is \emph{rational} if the range of all its
weight functions is in $\Qbot$. When we discuss algorithmic aspects of
weighted structures, we usually restrict our attention to rational
structures. We fix a reasonable encoding of rational weighted
structures by binary strings and let $\|\CA\|$ denote the length of
the encoding of $\CA$. Moreover, the \emph{order} $|\CA|$ of $\CA$ is
the number of elements of its universe, that is, $|\CA|\coloneq|A|$.
Note that for a fixed vocabulary $\Upsilon$, the coding length
$\|\CA\|$ is polynomial in the order $|\CA|$ and the maximum bitlength
of the
weights. 

\subsection{Feedforward Neural Networks}
\label{sec:fnn}
We assume that the reader is, at least superficially, familiar with
neural networks. We consider only the simplest type of neural network,
the feedforward neural network, also known as a multilayer
perceptron. We view neural networks as directed acyclic graphs with
weighted edges and nodes. Computation takes place at the nodes, or
\emph{neurons}. They form an affine linear combination of the values
at their in-neighbours, and then apply a (typically nonlinear)
\emph{activation function}. The resulting value is passed to the
out-neighbours of the node. Nodes of in-degree $0$ are input nodes;
nodes of out-degree $0$ are output nodes. The affine linear
combination a node forms is determined by the weights of the incoming
edges and the weight of the node itself (called the \emph{bias} of
the node). The only activation functions we consider here are the \emph{rectified linear unit} defined by $\relu(x)\coloneqq\max\{0,x\}$ and the identity function $\id(x)=x$ (only at output nodes).

As the reader will know, in practice, the weights and biases of an FNN are learned from data consisting of argument-value pairs of the function to be represented by the neural network. We are not concerned with the learning process here, but only with analysing the learned model.

Formally, a \emph{feedforward neural network (FNN)} is a weighted structure $\CN$ of vocabulary $\Upsilon_{\textup{FNN}}\coloneqq\{\wt,\bias,\lein,\leout\}$ consisting of a binary weight-function symbol $\wt$, a unary weight-function symbol $\bias$, and two binary relation symbols $\lein,\leout$ satisfying the following conditions:
\begin{itemize}
\item $E^\CN\coloneqq\big\{(v,w)\in N^2\bigmid wt^\CN(v,w)\neq\bot\big\}$ is the edge set of a directed acyclic graph on $V^\CN\coloneqq N$;
\item $\bias^\CN(v)=\bot$ if and only if $v$ is is an \emph{input node}, that is, a node of in-degree $0$ in the directed acyclic graph $(V^\CN,E^\CN)$;
\item $\le^\CN_{\mathsf{in}}$ is a linear order on the set of all
  input nodes and undefined on all other nodes, and
  $\le^\CN_{\mathsf{out}}$ is a linear order on the set of all
  \emph{output nodes}, that is, nodes of out-degree $0$, and undefined on all other nodes.
\end{itemize}
The \emph{input dimension} of an FNN is the number of input nodes, and the \emph{output dimension} is the number of output nodes.

An FNN $\CN$ of input dimension $m$ and output dimension $n$ defines a function $f^\CN\colon\Real^m\to\Real^n$. To define this function, we inductively define a function $f_v^\CN: \Real^m\to\Real$ for every node $v\in V^\CN$. If $v$ is the $i$th input node with respect to the linear order $\lein$, then we let $f_v^\CN(x_1,\ldots,x_m)\coloneqq x_i$. If $v$ has in-neighbours $u_1,\ldots,u_k$, we let\footnote{Our way of defining the function computed by a node is non-standard in that we do not yet apply the activation function, but only apply the $\relu$-activation when we feed the value into the next node. This way, we ensure that at the output nodes we apply no activation, or equivalently, the identity activation. We want to do this because, if we apply $\relu$ to all output values, our neural networks can only compute nonnegative functions.}
\[
f_v^\CN(x_1,\ldots,x_m)\coloneqq\bias^\CN(v)+\sum_{i=1}^k\wt(u_i,v)\cdot\relu\big(f_{u_i}^\CN(x_1,\ldots,x_m)\big).
\]
Finally, if $v_1,\ldots,v_n$ are the output nodes of $\CN$ listed in the order $\leout^\CN$ then we let
\[
  f^\CN(x_1,\ldots,x_m)\coloneqq\big(f^\CN_{v_1}(x_1,\ldots,x_m),\ldots,
  f^\CN_{v_n}(x_1,\ldots,x_m)\big).
\]

Let me remark that we can represent other, more complex neural networks as weighted structures in a similar way. But we restrict our attention to FNNs in this paper.

\section{First-Order Logic with Weight Aggregation}
\label{sec:fosum}
Our base logic is a straightforward extension of first-order logic to weighted structures. Variables and quantification range over the (finite) universe of a structure, and the numerical sort only provides values for terms. \emph{First-order logic $\FO$ over weighted structures} has two kinds of syntactical objects, \emph{formulas} $\phi$ and \emph{weight terms} $\theta$ (just called \emph{terms} in the following), which are defined by the following grammar:
\begin{align}
  \label{eq:fo1}
  \phi&::=x=y \mid R(x_1,\ldots,x_{\ar(R)})\mid\theta\le\theta\mid\neg\phi\mid(\phi*\phi)\mid
        Qx\phi\\
  \label{eq:fo2}
  \theta&::=0\mid 1\mid F(x_1,\ldots,x_{\ar(F)})\mid (\theta\circ\theta).
\end{align}
Here $x,y,x_i$ are variables, $R$ is a relation symbol,
$*\in\{\vee,\wedge,\to\}$ is a Boolean connective,
$Q\in\{\exists,\forall\}$ is a quantifier, $F$ is a
weight-function symbol, and $\circ\in\{+,-,\cdot,/\}$ is an arithmetic
operator. An \emph{$\FO$ expression} is either a formula or a term.
The \emph{vocabulary} of an expression $\xi$ is the set of all
relation and weight function symbols appearing in $\xi$. (Note that the relation $\le$, the
constants $0,1$, and the arithmetic operations $+,-,\cdot,/$ are not part of the vocabulary; we think of them as ``built-in'' relations, functions, constants.) We define the \emph{free variables} of an expression in the
obvious way. A \emph{closed expression} is an expression without free
variables. Closed formulas are also called \emph{sentences}. We write
$\xi(x_1,\ldots,x_k)$ to denote that the free variables of an
expression $\xi$ are among $x_1,\ldots,x_k$. (Not all of these
variables actually have to appear in the expression.)

The semantics of $\FO$ over weighted structures is defined in the obvious way. We define a value $\sem{\xi}^\CA(a_1,\ldots,a_k)$ for every $\FO$ expression $\xi(x_1,\ldots,x_k)$, every weighted structure $\CA$, and all elements $a_1,\ldots,a_k\in A$. If $\xi$ is a formula then $\sem{\xi}^\CA(a_1,\ldots,a_k)$ is a Boolean value ($0$ or $1$), and if  $\xi$ is a term then $\sem{\xi}^\CA(a_1,\ldots,a_k)\in\Rbot$. If the vocabulary of $\xi$ is not contained in the vocabulary of $\CA$ we let $\sem{\xi}^\CA(a_1,\ldots,a_k)\coloneqq0$ if $\xi$ is a formula and $\sem{\xi}^\CA(a_1,\ldots,a_k)\coloneqq\bot$ if $\xi$ is a term.
Otherwise, the value $\sem{\xi}^\CA(a_1,\ldots,a_k)$ is defined by a straightforward induction, using the usual semantics of the Boolean connectives and quantifiers and the rules of arithmetic over the extended reals $\Rbot$ explained at the beginning of Section~\ref{sec:prel}.

For closed expressions $\xi$, we write $\sem{\xi}^\CA$ instead of $\sem{\xi}^\CA()$, and for formulas $\phi(x_1,\ldots,x_k)$ we write $\CA\models\phi(a_1,\ldots,a_k)$ instead of $\sem{\xi}^\CA(a_1,\ldots,a_k)=1$, or just $\CA\models\phi$ if $\phi$ is a sentence.

Observe that for every rational $q\in\Qbot$ there is a closed term $\theta_q$ with empty vocabulary such that $\sem{\theta_q}^\CA=q$ for all structures $\CA$ (we let $\theta_\bot\coloneqq 1/0$). This means that we can actually use all elements of $\Qbot$ as constants in $\FO$ expressions. It is easy to see that the $\FO$ terms express precisely all rational functions (quotients of polynomials) with rational coefficients in the weights of a structure.

\begin{example}\label{exa:weighted-graph2}
  \renewcommand{\triangle}{\logic{triangle}}
  We can define the edge set $E^\CG$ of a weighted graph $\CG$ (see
  Example~\ref{exa:weighted-graph1}) by the formula
  $\edge(x,y)\coloneqq\wt(x,y)\neq\bot$, which is an abbreviation for $\neg(\wt(x,y)\le\bot\wedge \bot\le\wt(x,y))$.\footnote{From now on, we use similar abbreviations without commenting on it. We also omit unnecessary parentheses to improve readability.}
    Then the formula
  $\triangle(x,y,z)\coloneqq
  \edge(x,y)\wedge\edge(y,z)\wedge\edge(z,x)$
  defines the set of all triangles in $\CG$.

  The following formula defines the set of all triangles of minimum weight.
  \begin{multline*}
    \logic{min-wt-triangle}(x,y,z)\coloneqq
    \triangle(x,y,z)\wedge\forall x'\forall y'\forall z'\big(\triangle(x',y',z')\\
    \to\wt(x,y)+\wt(y,z)+\wt(z,x)\le \wt(x',y')+\wt(y',z')+\wt(z',x')\big).
  \end{multline*}
  \uend
\end{example}

The expressiveness of $\FO$ over weighted structures is very limited.
For example, it is easy to prove that there is no $\FO$ formula expressing that the sum
of the weights of the incoming edges to a node equals the sum of the
weights of the outgoing edges. To prove such a result, consider weighted
graphs where all edge weights are $1$. Then local isomorphisms between
tuples of vertices of such weighted graphs are just local isomorphisms
between the tuples in the underlying unweighted graphs, and we can use
standard Ehrenfeucht-\Fraisse-game arguments. Similarly, it can be
shown that there is no term expressing the
affine linear function needed to evaluate a neural network (see
Section~\ref{sec:fnn}). 

\subsection{\FOSUM}

To obtain a more expressive logic, we add an aggregation operator that
allows us to take sums over definable sets. We also add
a conditional operation, which is convenient, but can be
viewed as ``syntactic sugar'' (see Example~\ref{exa:ite}). We define
the formulas and terms of the logic \FOSUM\ using the rules
\eqref{eq:fo1} and \eqref{eq:fo2} and adding two new term-formation
rules:
\begin{equation}
  \label{eq:fosum}
  \theta::=\ite{\phi}{\theta}{\theta}
  \mid\sum_{(x_1,\ldots,x_k):\phi}\theta.
\end{equation}
The summation operator $\sum_{(x_1,\ldots,x_k):\phi}$ binds the
variables $x_1,\ldots,x_k$. Hence, the free variables of
$\sum_{(x_1,\ldots,x_k):\phi}\theta$ are those of $\phi$ and $\theta$
except $x_1,\ldots,x_k$. The free variables of
$\ite{\phi}{\theta_1}{\theta_2}$ are those of $\phi,\theta_1,\theta_2$.

The semantics of the conditional term $\ite{\phi}{\theta_1}{\theta_2}$
is obvious: if $\phi$ holds it takes the value of $\theta_1$ and
otherwise the value of $\theta_2$. To define the semantics of the
summation operator, suppose that
$\phi=\phi(x_1,\ldots,x_k,y_1,\ldots,y_\ell)$ and
$\theta=\theta(x_1,\ldots,x_k,y_1,\ldots,y_\ell)$. Then for the term
$\Big(\sum_{(x_1,\ldots,x_k):\phi}\theta\Big)(y_1,\ldots,y_\ell)$, a
structure $\CA$, and elements $b_1,\ldots,b_\ell\in A$ we
let
\[
  \sem{\sum_{(x_1,\ldots,x_k):\phi}\theta}^\CA(b_1,\ldots,b_\ell)
  \coloneqq\sum_{
    \substack{(a_1,\ldots,a_k)\in A^k\\
      \CA\models\phi(a_1,\ldots,a_k,b_1,\ldots,b_\ell)}}
  \sem{\theta}^\CA(a_1,\ldots,a_k,b_1,\ldots,b_\ell).
\]
We define the sum over the empty set to be $0$. Furthermore, if one of
the summands is undefined ($\bot$), then the whole sum is undefined.

\begin{example}\label{exa:weighted-graph3}
  \renewcommand{\triangle}{\logic{triangle}}
  Let us start by considering weighted graphs again. Of course the \FO\ formulas
  $\edge(x,y)$ and $\triangle(x,y,z)$ defined in
  Example~\ref{exa:weighted-graph2} are \FOSUM\ formulas as well. Then
  the terms
  \[
    \logic{\#edges}\coloneqq\sum_{(x,y):\edge(x,y)}1
    \qquad\text{and}\qquad
    \logic{\#triangles}\coloneqq\sum_{(x,y,z):\triangle(x,y,z)}1
  \]
  define the numbers of edges and triangles, respectively.
\uend
\end{example}

\begin{example}\label{exa:num-weights}
  Next, we turn to FNNs. As in Example~\ref{exa:weighted-graph2}, we
  let $\edge(x,y)\coloneqq\wt(x,y)\neq\bot$ be a formula defining the
  edge relation. Then the term
  \[
    \logic{\#weights}\coloneqq\sum_{(x,y):\edge(x,y)}1+\sum_{x:\bias(x)\neq\bot}1
  \]
  defines the total number of weights of an FNN.
  \uend
\end{example}

\begin{example}\label{exa:eval-fnn}
  For every $d\in\Nat$, we shall construct an \FOSUM\ term $\eval_d(x)$
  evaluating the function computed at a node $x$ of depth at most $d$
  in an FNN. Here, the  \emph{depth} of a node $v$ in an FNN $\CN$ is the length of the
  longest path from an input node to $v$ in the directed acyclic graph
  $(V^\CN,E^\CN)$. The notation we use in this example is explained in Section~\ref{sec:fnn}.

  If we want to evaluate neural networks, we have to specify an input.
  We do this by adding a unary weight function $\inp$ to the vocabulary, letting
  $\Upsilon_{\textup{FNN}^I}\coloneqq
  \Upsilon_{\textup{FNN}}\cup\{\inp\}$. Then an \emph{FNN with input}
  is an $\Upsilon_{\textup{FNN}^I}$-structure $\CN^I$ such that the
  $\Upsilon_{\textup{FNN}}$-reduct $\CN$ of $\CN^I$ is an FNN
  and $\inp^{\CN^I}(v)\neq\bot$ if and only if $v$ is an input node of
  $\CN$. For an FNN $\CN$ of input dimension $m$, say, with input
  nodes $u_1,\ldots,u_m$ listed in the order $\lein$, and
  an input $(r_1,\ldots,r_m)\in\Real^m$ we let $\CN(r_1,\ldots,r_m)$
  be the FNN with input expanding $\CN$ with
  $\inp^{\CN(r_1,\ldots,r_m) }(u_i)=r_i$.

  By induction on $d\ge0$, we shall define an $\FOSUM$ term
  $\eval_d(x)$ such that for every $m\in\Nat$, every FNN $\CN$ of input dimension
  $m$, every node $v\in V^\CN$,  and every $(r_1,\ldots,r_m)\in\Real^m$ it holds that
  \[
    \sem{\eval_d}^{\CN(r_1,\ldots,r_m)}(v)=\begin{cases}
      f_v^\CN(r_1,\ldots,r_m)&\text{if the depth of $v$ is at most
        $d$},\\
      \bot&\text{otherwise}.
    \end{cases}
  \]
  We let $\eval_0(x)\coloneqq \inp(x)$ and
  \begin{multline*}
    \eval_{d+1}(x)\coloneqq
    \textsf{if }\inp(x)\neq\bot\textsf{ then }\inp(x)\\
    \textsf{else }\bias(x)+
      \sum_{y:\edge(y,x)}\wt(y,x)\cdot\ite{\eval_d(y)\ge
      0}{\eval_d(y)}{0\cdot\eval_d(y)}.
  \end{multline*}
    To understand this term, note that the subterm $\ite{\eval_d(y)\ge
      0}{\eval_d(y)}{0}$ computes $\relu$ of $\eval_d(y)$. In the else clause, we write  $0\cdot\eval_d(y)$ instead of just $0$ to make sure the term is undefined if $\eval_d(y)=\bot$.

    Using the term $\eval_{d}(x)$, for every $i\in\Nat$ we can define
    a closed term $\eval_{d,i}$ such that for all $m,n\in\Nat$, all
    FNNs $\CN$ of input dimension $m$ and output dimension $n$, and all
    inputs $(r_1,\ldots,r_m)\in\Real^m$, if the depth of $\CN$ is at
    most $d$ and $i\le n$ then $\sem{\eval_d}^{\CN(r_1,\ldots,r_m)}$
    is the $i$th coordinate of $f^\CN(r_1,\ldots,r_m)\in\Real^n$, and
    otherwise $\sem{\eval_d}^{\CN(r_1,\ldots,r_m)}=\bot$.
    \uend
\end{example}

The previous example shows that we can evaluate FNNs of bounded depth
in \FOSUM. It is not hard to prove that the bounded-depth assumption
cannot be dropped. This follows from Theorem~\ref{theo:fosum-fully-ma}
below. Before we get there, let us consider a few more examples.

\begin{example}\label{exa:aggregation}
  Summation is only one possible aggregation operator. Other common
  aggregations are arithmetic mean, count, minimum and maximum. All
  these are definable in $\FOSUM$ using just summation.
  \begin{enumerate}
  \item For tuples $\vec x=(x_1,\ldots,x_k),\vec y=(y_1,\ldots,y_\ell)$ of variables and a formula
    $\phi(\vec x,\vec y)$ we define a term
    $\logic{count}_{\vec x:\phi}(\vec y)$ by
    $
      \logic{count}_{\vec x:\phi}\coloneqq\sum_{\vec
        x:\phi}1.
      $
      
    Then for all structures $\CA$ and 
    $\vec b\in A^\ell$ we have
    \[
      \sem{\logic{count}_{\vec
          x:\phi}}^\CA(\vec b)=\big|\big\{\vec a\in
      A^k\bigmid
      \CA\models\phi(\vec a,\vec b)\big\}\big|.
    \]
  \item For tuples $\vec x=(x_1,\ldots,x_k),\vec y=(y_1,\ldots,y_\ell)$ of variables, a formula
    $\phi(\vec x,\vec y)$, and a term $\theta(\vec x,\vec y)$ we define a term
    $\Big(\avg_{\vec x:\phi}\theta\Big)(\vec y)$ by
    $
      \avg_{\vec x:\phi}\theta\coloneqq\Big(\sum_{\vec
      x:\phi}\theta\Big)/\logic{count}_{\vec
      x:\phi}.
    $
    
  Then for all structures $\CA$ and elements
    $\vec b\in A^\ell$ the value  $\sem{\avg_{\vec
          x:\phi}\theta}^\CA(\vec b)$ is the arithmetic
      mean of the values
      $\sem{\theta}^\CA(\vec a,\vec b)$ for all
      $\vec a\in A^k$ such that
      $\CA\models\phi(\vec a,\vec b)$. The
      arithmetic mean over the empty set is undefined, and indeed, if there are
      no $\vec a\in A^k$ such that
      $\CA\models\phi(\vec a,\vec b)$ then $\sem{\avg_{\vec
          x:\phi}\theta}^\CA(\vec b)=\bot$.
    \item For tuples $\vec x=(x_1,\ldots,x_k),\vec y=(y_1,\ldots,y_\ell)$ of variables, a formula
    $\phi(\vec x,\vec y)$, and a term $\theta(\vec x,\vec y)$ we let
    $
      \phi'(\vec x,\vec y)\coloneqq\phi(\vec x,\vec y)\wedge\forall\vec
      x'\big(\phi(\vec x',\vec y)\to\theta(\vec x',\vec y)\le\theta(\vec
      x,\vec y)\big).
      $
      
    We
    define a term
    $\Big(\logic{max}_{\vec x:\phi}\theta\Big)(\vec y)$ by
    $
      \logic{max}_{\vec x:\phi}\theta\coloneqq\avg_{\vec
      x:\phi'}\theta.
    $
    
    Then for all structures $\CA$ and 
    $\vec b\in A^\ell$ we have
    \[
      \sem{\logic{max}_{\vec
          x:\phi}\theta}^\CA(\vec b)=\max\Big\{\sem{\theta}^\CA(\vec
      a,\vec b)\Bigmid\vec a\in A^k\text{ such that }
      \CA\models\phi(\vec a,\vec b)\Big\}.
    \]
    Similarly, we can define a term $\logic{min}_{\vec x:\phi}\theta$.
    \uend
  \end{enumerate}
\end{example}

\begin{example}\label{exa:ite}
  In this example, we will show that we can express the conditional
  $\ite{\phi}{\theta_1}{\theta_2}$ using only $\FO$ and the summation
  operation. For simplicity, we assume that $\phi,\theta_1,\theta_2$
  are closed expressions; the generalisation to arbitrary expressions
  is straightforward.

  Note that we have not used the conditional in
  Example~\ref{exa:aggregation}, so we can use the aggregation operators
  defined there. Observe that all
  structures $\CA$ we have
  \[
    \sem{\logic{count}_{x:\phi}}^\CA=
    \begin{cases}
      |\CA|&\text{if }\CA\models\phi,\\
      0&\text{otherwise}.
    \end{cases}
  \]
  Let $\eta\coloneqq
  \big(\logic{count}_{x:\phi}\big)/\big(\logic{count}_{x:x=x}\big)$.
  Then $\sem{\eta}^\CA=1$ if $\CA\models\phi$ and $\sem{\eta}^\CA=0$
  otherwise.
  Thus the term $\eta\cdot\theta_1+ (1-\eta)\cdot\theta_2$ is
  equivalent to $\ite{\phi}{\theta_1}{\theta_2}$.
  \uend
\end{example}

\begin{example}\label{exa:useless-nodes}
  Trying to understand the inner workings of neural networks, we can
  look for edges that have no effect on the result of a computation
  for a given input. We call such edges \emph{useless}.

  Let $\eval_{d}(x)$ be the \FOSUM\ term (from Example~\ref{exa:eval-fnn})
  defining the value of an FNN with input at a node $x$ of depth at
  most $d$.
Let $\edge'(x,y,x_0,y_0)\coloneqq\edge(x,y)\wedge\neg(x=x_0\wedge
  y=y_0)$. This formula defines the edge relation of the FNN obtained
  by removing edge $(x_0,y_0)$. Let $\eval'_{d}(x,x_0,y_0)$ be the
  term obtained from $\eval_{d}(x)$ by replacing each subformula
  $\edge(x,y)$ by $\edge'(x,y,x_0,y_0)$. Then for every FNN $\CN$ of input dimension $m$, every input $\vec
  r=(r_1,\ldots,r_m)\in\Real^m$, and all $v,u_0,v_0\in V^\CN$ such
  that the depth of $v$ is at most $d$ we have
  \[
    \sem{\eval'_d}^{\CN(r_1,\ldots,r_m)}(v,u_0,v_0)
    =\sem{\eval_d}^{\big(\CN-(u_0,v_0)\big)(r_1,\ldots,r_m)}(v)=
    f^{\CN-(u_0,v_0)}_v(r_1,\ldots,r_m),
  \]
  where $\CN-(u_0,v_0)$ is the FNN obtained from $\CN$ by deleting the
  edge $(u_0,v_0)$, that is, setting $\wt^{\CN-(u_0,v_0)
  }(u_0,v_0)\coloneqq\bot$.

  Then the formula
  \[
    \logic{useless}(x_0,y_0)\coloneqq\edge(x_0,y_0)\wedge\forall x\big(x\leout
    x\to\eval_d(x)=\eval_d(x,x_0,y_0)\big).
  \]
  defines the set of all useless edges.
  \uend
\end{example}

Our last example shows that we can also define natural global
properties of FNNs that do not depend on a particular input. This
example is much more involved (it is actually a deep theorem proved in
\cite{GroheSSV25}), and we do not give any details.

\begin{example}[\cite{GroheSSV25}]\label{exa:integrate}
  In this example, we want to show that we can define integrals over
  functions computed by FNNs in \FOSUM. We need to fix the depth $d$
  and the input dimension $m$ of our FNNs.

  We specify the boundaries of the region we want to integrate over by
  $2m$ weight constants $\mathsf a_1,\ldots,\mathsf a_m,\mathsf
  b_1,\ldots,\mathsf b_m$ (recall that weight constants are $0$-ary
  weight functions). In this example, for an FNN $\CN$ and
  $\vec a=(a_1,\ldots,a_m),\vec b=(b_1,\ldots,b_m)\in\Real^m$ with $a_i\le b_i$ for all $i\in[m]$,
  by $\CN(\vec a,\vec b)$ we denote the expansion of
  $\CN$ to the vocabulary $\Upsilon_{\textup{FNN}}\cup\{\mathsf a_1,\ldots,\mathsf a_m,\mathsf
  b_1,\ldots,\mathsf b_m\}$ with $\mathsf a_i^{\CN(\vec a,\vec
    b)}=a_i$ and $\mathsf b_i^{\CN(\vec a,\vec b)}=b_i$.

  Then there is a closed \FOSUM\ term $\logic{integrate}_{d,m}$ such
  that for all FNNs $\CN$ of input dimension $m$, output dimension
  $1$, and depth at most $d$ and all $\vec a=(a_1,\ldots,a_m),\vec b=(b_1,\ldots,b_m)\in\Real^m$ with $a_i\le b_i$ we have
  \[
    \sem{\logic{integrate}_{d,m}}^{\CN(\vec a,\vec
      b)}=\int_{a_1}^{b_1}\cdots
    \int_{a_m}^{b_m}f^{\CN}(x_1,\ldots,x_m)dx_1\ldots dx_m.
  \]
  The proof is difficult. It is based on the \FOSUM\ definability of
  cylindrical cell decomposition of the (piecewise linear) functions
  computed by FNNs. To get at least an idea of how integrals can be
  defined, let us consider the special case $d=2,m=1$. That is, we look
  at FNNs $\CN$ with one input node, one hidden layer, and on output
  node. Such an FNN computes a piecewise linear function
  $f^\CN\colon\Real\to\Real$. The breakpoints of this function are determined
  by the nodes on the hidden layer of $\CN$, and using these nodes,
  we can define the coordinates of all the breakpoints. We can,
  moreover, define the slopes of the linear pieces connecting the
  breakpoints. From this, the integral is easy to compute.
  \uend
\end{example}

\subsection{Complexity}
In this section, we assume all weighted structures to be rational (see Section~\ref{sec:weighted-structures}).
It is a well-known fact that the evaluation of formulas of first-order
logic over (unweighted) finite relational structures is in the
complexity class uniform $\AC^0$, the class of problems decided by
dlogtime-uniform families of Boolean circuits over $\neg,\wedge,\vee$
of unbounded fan-in, bounded depth, and polynomial size (see
\cite{Immerman99}).

To evaluate \FO\ formulas and terms over rational
weighted
structures, we need to evaluate terms over the rationals. It is known
that this is possible in the complexity class uniform $\TC^0$,
defined like $\AC^0$ but allowing threshold gates in the circuits in
addition to the standard logic gates \cite{BarringtonIS90,HesseAB02}. As
iterated addition (addition of a family of $n$ numbers of bitlength
$n$) is in uniform $\TC^0$, we can actually carry out all the
arithmetic needed to evaluate $\FOSUM$ formulas and terms and hence
obtain the following theorem.\footnote{It is difficult to trace the
  results on arithmetic in uniform $\TC^0$. It was proved in the 1980s
  (going back to \cite{ChandraSV84}) that arithmetic is in non-uniform
  $\TC^0$, but for addition, subtraction, iterated addition, and
  multiplication, it was known that the constructions could actually be
  made uniform. Certainly, this was known at the time of Barrington,
  Immerman and Straubing's work on logic and uniform $\TC^0$
  \cite{BarringtonIS90}. However, I am not aware of a reference where these results are proved. Iterated addition is critical; therefore, I
  added a proof that iterated addition is in uniform $\TC^0$ to
  my recent paper \cite{Grohe24b}. The most difficult case is division; it was
  only proved by Hesse in 2000 \cite{Hesse01,HesseAB02} that division
  is in uniform $\TC^0$.}

\begin{theorem}\label{theo:tc0}
  The data complexity of $\FOSUM$ is in uniform $\TC^0$.

  That is, for
  every closed $\FOSUM$ expression $\xi$ the value $\sem{\xi}^\CA$ of
  $\xi$ in a given structure $\CA$ can be computed by a dlogtime
  uniform family of threshold circuits of bounded depth and
  polynomial size.
\end{theorem}

\subsection{Model-Agnostic Queries}
Often, we are only interested in the function a neural network
computes and not in the structure of the network. Note that for every function
computable by an FNN, there are infinitely many very different FNNs
computing this function. We call queries whose outcome only depends on
the function and not on the specific FNN \emph{model agnostic}.

A closed \FOSUM\ expression $\xi$ is \emph{model agnostic on a class $\BN$} of FNNs if for all $\CN,\CN'\in\BN$, if
$f^{\CN}=f^{\CN'}$ then $\sem{\xi}^{\CN}=\sem{\xi}^{\CN'}$. We can extend this
definition to other logics on weighted structures, in particular to the
logic \IFPSUM\ that will be discussed in Section~\ref{sec:ifpsum}. We
can even generalise the definition to abstract \emph{Boolean queries} on $\BN$,
that is, isomorphism-invariant functions $\Phi\colon\BN\to\{0,1\}$, and to
isomorphism invariant functions $\Theta\colon\BN\to\Rbot$. In a different
direction, we can also generalise it to classes of expansions of
FNNs such as FNNs with input (see Example~\ref{exa:eval-fnn}) and FNNs with
additional weight constants (see Example~\ref{exa:integrate}). But note that
the invariance condition $f^{\CN}=f^{\CN'}$ always refers to the plain
neural networks. For example, for neural networks with input $\CN(\vec
r),\CN'(\vec r')$ we still require $f^{\CN}=f^{\CN'}$ and not
$f^{\CN}(\vec r)=f^{\CN'}(\vec r')$.

We denote the class of all FNNs of input dimension $m$ and output
dimension $n$ by $\BN(m,n)$. Moreover, we let
$\BN(m,*)\coloneqq\bigcup_{n\ge 1}\BN(m,n)$ and define $\BN(*,n)$ and
$\BN(*,*)$ similarly. Note that $\BN(*,*)$ is the class of all FNNs. We
denote the class of all FNNs of depth at most $d$ by $\BN_d(*,*)$, and
for $m,n\in\PNat\cup\{*\}$ we let
$\BN_d(m,n)\coloneqq\BN(m,n)\cap\BN_d(*,*)$. Moreover, we denote the
corresponding classes of FNNs with input by $\BN^I(m,n)$ and
$\BN_d^I(m,n)$.

We call an expression, query, or function \emph{fully model agnostic}
if it is model agnostic on the class of all FNNs or
expansions of FNNs of the
suitable form. We leave the ``suitable form'' and hence the class of
structures vague on purpose to be able to phrase results in the most
natural form. Typically, we consider the classes $\BN(*,*)$ or
$\BN(*,1)$, or $\BN^I(*,*)$ or $\BN^I(*,1)$ for FNNs with input.

\begin{example}\label{exa:ma1}
  The \FOSUM\ term $\eval_{d,i}$ of Example~\ref{exa:eval-fnn} is model
  agnostic on $\BN^I_d(*,*)$, but 
  it is not model agnostic on $\BN^I_{d+1}(1,1)$ and hence not fully
  model agnostic.

  However, for each $i\ge 1$ the
  evaluation function $\Eval_i\colon\BN^I(*,*)\to\Real$, defined by
  letting $\Eval_i(\CN(\vec r))$ be the $i$th entry of the vector $f^\CN(\vec r)$, is fully model agnostic. In the following, most often we restrict our
  attention to FNNs with a single output, for which we denote the
  evaluation function by $\Eval$ instead of $\Eval_1$.
\end{example}

\begin{example}\label{exa:not-ma}
  The term $\logic{\#weights}$ of Example~\ref{exa:num-weights} is not
  model-agnostic, because FNNs computing the same function may have
  different weights. Similarly, the sentence $\logic{\#weights}\le
  10.000$ is not model agnostic.

  However, for every function $f\colon\Real\to\Real$ the query ``Is there
  an FNN $\CN'$ with at most $10.000$ weights that computes the same
  function?'', that is, $\Phi\colon\BN(*,*)\to\{0,1\}$ defined by
  \[
    \Phi(\CN)\coloneqq
    \begin{cases}
      1&\text{there exists an $\CN'\in\BN(*,*)$ such that
        $\sem{\logic{\#weights}}^{\CN'}\le10.000$}\\
      &\hspace{1cm}\text{and $f^{\CN'}=f^{\CN}$},\\
      0&\text{otherwise}
    \end{cases}
  \]
 is fully model-agnostic.
  \uend
\end{example}

\begin{example}\label{exa:ma2}
  The  \FOSUM\ term $\logic{integrate}_{d,m}$ of Example~\ref{exa:integrate} is model
  agnostic on the class of suitable expansions of $\BN_d(m,1)$ by
  weight constants, but not fully model-agnostic.

  And again, we can define a fully model-agnostic integration function
  $\Integrate$ on the class of all FNNs with output dimension $1$
  expanded by suitable weight constants.

  We can also consider the special case $\Integrate_{[0,1]}$
  of the \Integrate\ query where we set all integration boundaries $a_i$
  to $0$ and all $b_i$ to $1$. Then there is no need to add explicit
  weight constants for the integration boundaries, and we can view
  $\Integrate_{[0,1]}$ as a fully model agnostic function on $\BN(*,1)$.
  \uend
\end{example}

\begin{example}\label{exa:zero}
  The Boolean query $\Zero$ on $\BN(*,*)$ defined by
  \[
    \Zero(\CN)\coloneqq
    \begin{cases}
      1&\text{if }f^{\CN}(\vec r)=0\text{ for all }\vec r\in\Real^m,\text{ where $m$ is the input dimension of $\CN$},\\
      0&\text{otherwise}
    \end{cases}
  \]
  is fully model-agnostic.

  It was proved in \cite{GroheSSV26a} that it is \NP-complete to
decide if $\Zero(\CN)=0$ for a rational FNN $\CN\in\BN(1,1)$.
(Wurm~\cite{Wurm24} had proved earlier that the same query on $\BN(*,1)$ is \NP-complete.)  
Hence, the $\NonZero$ query is NP-complete, which is unsurprising,
because it may be viewed as a form of satisfiability query for
FNNs. Let me remark that the proof of \cite{GroheSSV26a} shows that
the query $\NonZero_{[0,1]}$ on $\BN(1,1)$, asking
if there is an input $r\in[0,1]$ such that $f^\CN(r)\neq0$ for an
$\CN\in\BN(1,1)$, is NP-hard already.
This also implies that it is NP-hard to decide if the function
$\Integrate_{[0,1]}$ (see Example~\ref{exa:ma2}) is nonzero.
  \uend
\end{example}

We would like to understand which fully model-agnostic queries are
expressible in \FOSUM. The answer is disappointing: only the trivial
constant queries are.

\begin{theorem}[\cite{GroheSSV25}]\label{theo:fosum-fully-ma}
  Let $m,n\in\PNat$, and let $\phi$ be an \FOSUM\ sentence that is model agnostic on $\BN(m,n)$. Then either
  $\CN\models\phi$ for all $\CN\in\BN(m,n)$ or $\CN\not\models\phi$
  for all $\CN\in\BN(m,n)$.
\end{theorem}

A similar result holds for the class $\BN^I(m,n)$ of FNNs with input.
Hence the evaluation function $\Eval$ (see Example~\ref{exa:ma1}) is not
expressible in $\FOSUM$.

The proof of Theorem~\ref{theo:fosum-fully-ma} is easy. Intuitively,
it is based on the fact that $\FOSUM$ can only express local
properties, but we can extend edges in an FNN by long paths of
weight-1 edges without changing the function that is computed.

However, \FOSUM\ can express interesting model-agnostic queries and
functions on classes of FNNs of bounded depth (see
Examples~\ref{exa:ma1} and \ref{exa:ma2}).
As a benchmark for the expressivity of model-agnostic queries, we take
a logic that only has ``black box'' access to the function computed by
an FNN, but not to the FNN itself. The logic we consider is
first-order logic over the ordered field of the reals expanded by the
function computed by our FNN. For an $m$-ary function symbol $f$ we
let $\FO(\CR,f)$ be first-order logic in the language
$\{+,\cdot,0,1,\le,f\}$, where $+,\cdot$ are binary function symbols,
$0,1$ are constant symbols, and $\le$ is a binary relation symbol. We
interpret $\FO(\CR,f)$ formulas over expansions
$(\CR,\hat f)$ of the ordered field of the reals,
$\CR=(\Real,+,\cdot,0,1,\le)$ by an $m$-ary function $\hat
f\colon\Real^m\to\Real$. Every $\FO(\CR,f)$ sentence $\phi$ defines a query
over $\BN(m,1)$ by
\[
  \phi(\CN)=1\iff (\CR,f^\CN)\models\phi.
\]
No confusion should arise from the fact that we use the same symbol
$\phi$ for the query and the sentence that defines it. Clearly, this query is invariant over $\BN(m,1)$. Note that we need to
fix the input dimension and the output dimension to work in this
logic, because function symbols have fixed arity. It is not crucial
that the output dimension is $1$. If we wanted to query FNNs with
a (fixed) output dimension $n$, we could simply add $n$ function symbols
instead of just one.

We can also define functions from FNNs to the reals. Each
$\FO(\CR,f)$ formula $\phi(y)$ defines a function from $\BN(m,1)$ to $\Rbot$ by
\[
  \phi(\CN)=
  \begin{cases}
    s&\text{if $(\CR,f^\CN)\models\phi(s)$ and there is no $s'\neq s$
      such that $(\CR,f^\CN)\models\phi(s')$},\\
    \bot&\text{otherwise}.
  \end{cases}
\]
To define queries over FNNs with input, we use formulas
$\phi(x_1,\ldots,x_m)$ with $m$ free variables. Each such formula
defines a query over $\BN^I(m,1)$ by
\[
  \phi(\CN,r_1,\ldots,r_m)=1\quad:\Longleftrightarrow\quad (\CR,f^\CN)\models\phi(r_1,\ldots,r_m).
\]
Similarly, we can define queries defined over FNNs in $\BN(m,1)$ expanded by any
fixed number $\ell$ of weight constants using $\FO(\CR,f)$ formulas with $\ell$
free variables, and of course, we can also define functions using one
additional free variable.

We also consider the \emph{linear fragment} $\FO(\CRlin,f)$ of
$\FO(\CR,f)$ consisting of all formulas in the language
$\{+,(q)_{q\in\CQ},\le\}$, in which we add a constant symbol for
every rational. We interpret these formulas $\FO(\CR,f)$ formulas $\phi$ over expansions
$(\CRlin,\hat f)$ of 
$\CRlin=(\Real,+, (q)_{q\in\CQ},\le)$. It is useful to have
constants for the rationals in this language. As each rational number
is definable in $\CR$, we do not need these additional constants in the
logic $\FO(\CR,f)$. The formulas of $\FO(\CRlin,f)$ define queries
over FNNs in the same way as the formulas of $\FO(\CR,f)$.

\begin{example}
  The evaluation query $\Eval$ (see Example~\ref{exa:ma1}) on $\BN^I(m,1)$ is
  expressible in $\FO(\CRlin,f)$ by the formula
  $\phi(x_1,\ldots,x_m,y)\coloneqq f(x_1,\ldots,x_m)=y$.
\end{example}

\begin{example}\label{exa:zero2}
  The query $\Zero$ (see Example~\ref{exa:zero}) on $\BN(m,1)$ is
  expressible in $\FO(\CRlin,f)$ by the sentence
  $\phi\coloneqq\forall x_1\ldots\forall x_m f(x_1,\ldots,x_m)= 0$. 
\end{example}

\begin{example}
  We may want to understand how much a function computed by an FNN
  depends on a particular input feature. For an $\epsilon>0$, we call
  a coordinate $i\in[m]$ \emph{$\epsilon$-irrelevant} for a function
  $f\colon\Real^m\to\Real$ if for all $r_1,\ldots,r_m,r_i'\in \Real$
  it holds that
  \[
    |f(r_1,\ldots,r_m)-f(r_1,\ldots,r_{i-1},r_i',r_{i+1},\ldots,r_m)|\le\epsilon.
  \]
  For every
  (rational) $\epsilon>0$, we can easily define an
  $\FO(\CRlin,f)$-formula expressing that $i$ is $\epsilon$-irrelevant
  for $f$.

  Hence the query $\Irr_{\epsilon,i}$ on $\BN(m,1)$ defined by
  $\Irr_{\epsilon,i}(\CN)=1$ if $i$ is $\epsilon$-irrelevant
  for $f^\CN$ is expressible in $\FO(\CRlin,f)$.
  \uend
\end{example}

By Theorem~\ref{theo:fosum-fully-ma}, the queries $\Zero$ and
$\Irr_{\epsilon,i}$ on $\BN(1,1)$ are not expressible in $\FOSUM$. It is
not even obvious how to express the queries on FNNs of bounded depth.
However, the following theorem shows that for all $d,m$ they are \FOSUM-expressible on $\BN_d(m,1)$.

\begin{theorem}[\cite{GroheSSV25}]\label{theo:black2white}
  Let $d\in\PNat$. Then every query on $\BN_d(m,1)$ that is
  expressible in $\FO(\CRlin,f)$ is expressible in $\FOSUM$.
\end{theorem}

In fact, the theorem also extends to FNNs with input and to arbitrary
expansions by weight functions. The proof is difficult. The problem is
that $\FO(\CRlin,f)$ can quantify over arbitrary reals, whereas
$\FOSUM$ formulas only have access to the finite set of weights in the
FNN. To resolve this issue, in \FOSUM\ we define a cylindrical cell
decomposition of the piecewise-linear function computed by our FNN.
Each of the cells can be dealt with uniformly, because the function is
linear in each cell. Then we can simulate quantification over the reals
by quantification over the cells.

\begin{example}\label{exa:zero3}
  Let us once more revisit the $\Zero$ query of
  Examples~\ref{exa:zero}, \ref{exa:zero2}. 
  It follows from Theorem~\ref{theo:black2white} that for each fixed
$m$ and $d$ the restrictions of $\Zero$ and $\NonZero$ to $\BN_d(m,1)$
are expressible in $\FOSUM$. Hence by Theorem~\ref{theo:tc0}, the
queries can be evaluated in uniform $\TC^0$. This is remarkable because, as pointed out,
$\NonZero$ is a form of satisfiability, and satisfiability for Boolean
circuits is already hard for circuits of depth $3$.
\uend
\end{example}

We close this section with an example showing that there are
model-agnostic queries on bounded-depth FNNs that are
expressible in $\FOSUM$, but not in $\FO(\CR,f)$.

\begin{example}[\cite{GroheSSV25}]
  The query $\int_0^1f^{\CN}(x)dx=0$ on $\BN_2(1,1)$ is expressible
  in \FOSUM\ (see Example~\ref{exa:integrate}), but not in
  $\FO(\CR,f)$. This follows from the fact that every continuous
  piecewise linear function $f\colon\Real\to\Real$ can be computed by an
  FNN of depth $2$ and \cite[Theorem~6.1]{GroheSSV25} stating that integration over continuous piecewise linear functions $f\colon\Real\to\Real$ is not expressible in $\FO(\CR,f)$. The proof is
  based on the (deep) fact that it is not expressible in $\FO(\CR,X)$ if a
  set $X\subseteq\Real$ has even cardinality (see \cite{KuperLP00,Libkin04}) and a (straightforward) reduction from even cardinality to integration.
  \uend
\end{example}

\section{Fixed-Point Logic with Weight Aggregation}
\label{sec:ifpsum}
While we may argue that in practice, we often have fixed neural
network architectures and just adapt the weights, the fact that we
cannot even express the evaluation function for FNNs of unbounded
depth in \FOSUM\ seriously limits its applicability as a query
language for FNNs.

\subsection{\IFPSUM}

To evaluate FNNs of unbounded depth, we need some form of
recursion. \emph{Fixed-point logic} is the classical mechanism for
formalising inductive definability in a first-order logic framework
(see \cite{EbbinghausF99,Moschovakis74}). We define the logic
\IFPSUM, \emph{inflationary fixed-point logic with weight
  aggregation}, following
\cite{GroheSSV26a}. The logic is closely related to the functional
fixed-point logic for $\Real$-algebras from \cite{GradelM95}. The formulas and weight terms of \IFPSUM\
are defined by the same rules as those for \FOSUM, \eqref{eq:fo1},
\eqref{eq:fo2}, and \eqref{eq:fosum} adding one additional
term-formation rule:
\begin{equation}
  \label{eq:ifpsum}
  \theta::=\ifp\big(
  F(x_1,\ldots,x_{\ar(F)})\gets\theta\big)(x'_1,\ldots,x'_{\ar(F)}),
\end{equation}
where the $x_i,x_i'$ are variables and $F$ is a weight function symbol. Note that the tuples
$(x_1,\ldots, x_{\ar(F)})$ and $(x'_1,\ldots,x'_{\ar(F)})$ are not
necessarily disjoint; in fact, we will typically let $x_i=x_i'$ for all $i$.

The \ifp-operator in \eqref{eq:ifpsum} binds the variables
$x_1,\ldots, x_{\ar(F)}$. Thus the free variables of the fixed-point
term $\ifp\big( F(\vec x)\gets\theta\big)(\vec x')$ are those of
$\theta$ minus the variables in $\vec x$ plus the variables in
$\vec x'$. The \ifp\ operator also binds the weight-function symbol
$F$; we say that $F$ is an \emph{intensional symbol} of the
fixed-point term. All other relation and weight-function symbols
appearing in $\theta$ are \emph{extensional symbols} of the
fixed-point term.

\IFPSUM\ faithfully extends \FOSUM, so 
to define the semantics of \IFPSUM, we only need to define the
semantics of the $\ifp$-operator. Consider a term
\begin{equation}
  \label{eq:2}
  \eta(\vec x',\vec y)\coloneqq \ifp\big(
  F(\vec x)\gets\theta\big)(\vec x'),
\end{equation}
where $\theta=\theta(\vec x,\vec y)$ is a term of vocabulary
$\Upsilon\cup\{F\}$. Thus, all extensional symbols of $\eta$ are in
$\Upsilon$. Let $k\coloneqq|\vec x|=|\vec x'|=\ar(F)$ and
$\ell\coloneqq|\vec y|$. Let $\CA$ be an $\Upsilon$-structure and
$\vec b\in A^\ell$. For every function $\hat F\colon A^k\to\Rbot$, let
$(\CA,\hat F)$ be the $\Upsilon\cup\{F\}$-expansion of $\CA$ with
$F^{(\CA,\hat F)}=\hat F$. We define a sequence $(F^{(i)})_{i\ge 0}$
of functions $F^{(i)}\colon A^k\to\Rbot$ as follows:
\begin{itemize}
\item $F^{(0)}(\vec
  a)\coloneqq\bot$ for all $\vec a\in A^k$;
\item
  $
  F^{(i+1)}(\vec a)\coloneqq
  \begin{cases}
    \sem{\theta}^{(\CA,F^{(i)})}(\vec a,\vec b)&\text{if }F^{(i)}(\vec a)=\bot,\\
    F^{(i)}(\vec a)&\text{if }F^{(i)}(\vec a)\neq\bot.
  \end{cases}
  $
\end{itemize}
Since we never change the value of $F^{(i)}(\vec a)$ once it is
defined ($\neq\bot$), there is an $i<|A|^k$ such that
$F^{(i)}=F^{(i+1)}$. Let $i_\infty$ be the least such $i$. Then
$F^{(i)}=F^{(i_\infty)}$ for all $i\ge i_\infty$.

Finally, the value of the term $\eta(\vec x',\vec y)$ in
\eqref{eq:2} is defined by
\[
  \sem{\eta}^{\CA}(\vec a',\vec b)\coloneqq F^{(i_\infty)}(\vec a').
\]

\begin{example}\label{exa:eval-ifp}
  The evaluation function $\Eval\colon\BN^I(*,*)\to\Real$ (see
  Example~\eqref{exa:ma1}) is expressible in \IFPSUM. We first define
  a term evaluating the function computed at every node of an FNN:
  \begin{multline*}
    \logic{eval-node}(x)\coloneqq\ifp\Big(F(x)\gets\textsf{if }\inp(x)\neq\bot\textsf{ then
                        }\inp(x)\\
      \textsf{else }\bias(x)+
      \sum_{y:\edge(y,x)}\wt(y,x)\cdot\ite{F(y)\ge
      0}{F(y)}{0\cdot F(y)}\Big)(x)
  \end{multline*}
Note how this term formalises the (meta-level) inductive definition of the terms
  $\eval_d(x)$ in Example~\ref{exa:eval-fnn} within the logic. It is
  easy to see that for every $m\in\PNat$, every FNN $\CN\in\BN(m,*)$, every
  input $\vec r\in\Real^m$, and every node $v$ we have
  $\sem{\logic{eval-node}}^{\CN(\vec r)}(v)=f^{\CN}_v(\vec r)$.

  Now we let
  $
    \logic{eval}\coloneqq\avg_{x:x\leout x}\logic{eval-node}(x).
  $
  Then for every FNN $\CN\in\BN(m,1)$ and every
  input $\vec r\in\Real^m$ we have
  $\sem{\logic{eval}}^{\CN(\vec r)}=f^{\CN}(\vec r)$.
  \uend
\end{example}

The definition of \IFPSUM\ allows nested fixed-point operators (the
term $\theta$ in \eqref{eq:ifpsum} may contain further fixed-point
operators), but the following theorem shows that these are not
necessary. A \emph{selection condition} is a conjunction of equalities
and inequalities between variables.

\begin{theorem}[\cite{GroheSSV26a}]
  \begin{enumerate}
    \item
     Every \IFPSUM\ formula $\varphi(\vec x)$ is equivalent to a formula of the form
       $
       \exists \vec y\big(\chi(\vec y)\wedge\ifp\big(F(\vec x,\vec y)\gets\theta\big)(\vec
       x,\vec y)\big) $,
     where $\chi$ is a selection condition and $\theta(\vec x,\vec y)$ is
     an \FOSUM\ term.
     \item
       Every \IFPSUM\ term $\eta(\vec x)$ is equivalent to a term of the form
       $\avg_{\vec y:\chi(\vec y)}\ifp\big(F(\vec x,\vec y)\gets\theta\big)(\vec
       x,\vec y)$, where again $\chi$ is a selection condition and $\theta(\vec x,\vec y)$ is
     an \FOSUM\ term. Furthermore, $\avg_{\vec y:\chi(\vec y)}$
     is the operator defined in Example~\ref{exa:aggregation}.
\end{enumerate}
\end{theorem}

The theorem is proved using standard techniques from finite model
theory (see \cite[Chapter~8]{EbbinghausF99}).

It is also possible to simulate simultaneous inductions with a single
fixed-point operator. In \cite{GroheSSV26a}, this was used to 
give a datalog-style syntax for \IFPSUM.

\subsection{Complexity}
An important property of (least or inflationary) fixed-point logic
over finite structures is that queries can be evaluated in polynomial
time and that, over ordered finite structures, all polynomial-time
queries can be expressed in the logic. We say that fixed-point logic
\emph{captures} polynomial time over ordered finite
structures. Similarly, it can be proved \cite{GradelM95} that over ordered weighted
structures, \IFPSUM\ captures
polynomial time in the Blum-Shub-Smale model of computation over the reals
\cite{BlumSS89,BlumCSS98}. However, here we are interested in
traditional bit complexity, and the following example shows that in
this model, \IFPSUM\ terms cannot be evaluated in polynomial time, even
over rational structures, because they can get too large.

\begin{example}[\cite{GradelG98}] \label{exa:squaring}
  Let
  \[
    \sigma(x)\coloneqq
    \ifp\Bigg(F(x)\gets
      \ite{\exists y\, \edge(y,x)}{\Big(\sum_{y:\edge(y,x)} F(y)\Big)\cdot
       \Big(\sum_{y:\edge(y,x)} F(y)\Big)}{2}\Bigg)(x).
 \]
 If we evaluate the term over an FNN $\CN\in\BN(1,1)$ that is just a
 path of length $d$, then, regardless of the weights, for the output
 node $v$ of $\CN$ it holds that $\sem{\sigma}^{\CN}(v)=2^{2^d}$.
 \uend
\end{example}

To resolve the issue with repeated squaring that the example shows, in \cite{GroheSSV26a} we introduced the \emph{scalar
  fragment} $\sIFPSUM$ of $\IFPSUM$ by restricting the use of
intensional weight function symbols in multiplications and divisions:
in an $\sIFPSUM$ expression $\xi$, for every subterm of the form
$\theta_1\cdot\theta_2$ either $\theta_1$ or $\theta_2$ contains no
intensional symbol, and for every subterm of the form
$\theta_1/\theta_2$ the term $\theta_2$ contains no intensional symbol. (The precise formal definition is a bit tricky,
because we need to specify which symbols count as intensional in
a specific context. I refer the reader to \cite{GroheSSV26a} for
details.)

\begin{example}
  The term $\eval$ of Example~\ref{exa:eval-ifp} is in \sIFPSUM. The
  term $\sigma(x)$ in Example~\ref{exa:squaring} is not.
  \uend
\end{example}

The following theorem shows that $\sIFPSUM$ expressions can be
evaluated in polynomial time over rational weighted structures.

\begin{theorem}[\cite{GroheSSV26a}]
  The data complexity of $\sIFPSUM$ is in \PTIME.
\end{theorem}

Let me remark that the proof of this theorem is not as trivial as it
may seem. The difficulty is to bound the size of the denominators of
the fractions appearing in the evaluation of an expression.

As for \FOSUM, we can now ask which model-agnostic queries over FNNs
we can express in \IFPSUM\ or \sIFPSUM.

\begin{example}
  If $\PTIME\neq\NP$, then the $\Zero$ query and the integration
  function $\Integrate_{[0,1]}$ are not expressible in
  $\sIFPSUM$. The reason is that $\NonZero$
  and $\Integrate_{[0,1]}\neq 0$ are $\NP$-hard (see
  Example~\ref{exa:zero}).
  \uend
\end{example}

While the reason for the inexpressibility in the previous example is
high computational complexity, it turns out that we cannot even
express all
queries computable in polynomial time, not even in \IFPSUM.

\begin{theorem}[\cite{GroheSSV26a}]\label{theo:not-ptime}
  There is a model-agnostic query on $\BN(1,1)$ that is
  decidable in polynomial time, but not expressible in \IFPSUM\ even
  on FNNs where all weights are $1$ or $0$.
\end{theorem}

To prove this theorem, we exploit the fact that \IFPSUM\ only has a
limited ability to carry out computations on the numerical part of the
weighted structure, because it has no variables ranging over numbers
and hence no quantification over numbers.

So \IFPSUM\ is limited in its ability to carry out computations with
rationals of large bit size. This turns out to be the main limitation
when it comes to polynomial-time computable queries. In
\cite{GroheSSV26a}, we define a normal form for FNNs, \emph{reduced
  FNNs}, which are obtained by factoring through an equivalence
relation similar to bisimilarity. Every FNN $\CN$ is equivalent to a unique
reduced FNN $\tilde\CN$. For every $b\in\Nat$, we let $\BR_b(*,*)$ be the class of
all FNNs $\CN$ such that all weights of the reduced FNN $\tilde\CN$
have bitsize at most $b$.

\begin{theorem}[\cite{GroheSSV26a}]
  For every $b\in\Nat$, every polynomial-time decidable model-agnostic query on
  $\BR_b(*,*)$ is expressible in $\sIFPSUM$.
\end{theorem}

\section{Concluding Remarks}

We studied two logics over weighted structures, $\FOSUM$ and $\IFPSUM$,
mainly with respect to their ability to express queries over
feedforward neural networks. While quite a few nontrivial
results about these logics have been proved in 
\cite{GroheSSV26a,GroheSSV25}, many interesting question remain open.

A central result is Theorem~\ref{theo:black2white}, stating that
\FOSUM\ can simulate $\FO(\CRlin,f)$ on bounded depth FNNs. Can we extend this
theorem from $\CRlin$ to $\CR$? Is there a way to lift the results to
FNNs of arbitrary input dimension? For this, we would have to adapt
the definition of $\FO(\CRlin,f)$, because as it is, the input
dimension of the FNNs we can consider is simply the arity of the fixed
function symbol $f$.

Our original hope was that we might be able to drop the bounded depth
assumption in Theorem~\ref{theo:black2white} if we replace $\FOSUM$ by
$\IFPSUM$. In view of the NP-completeness of simple
$\FO(\CRlin,f)$-expressible queries like $\NonZero$, this seems
unlikely. But can we actually prove that $\NonZero$
is not expressible in \sIFPSUM\ or \IFPSUM\ without making
a complexity-theoretic assumption?

We showed that other aggregation operators (counting, arithmetic mean,
minimum and maximum) can be expressed in terms of summation alone. In
general, aggregation functions can be defined as functions from
multisets of reals to the reals (see~\cite{GradelG98}). It would be
interesting to develop an understanding for the inter-definability
between different aggregation operators. For example, it seems
unlikely that summation can be defined in terms of arithmetic mean.

In a different direction, is there is a query language that
expresses precisely the model-agnostic queries on FNNs that are
polynomial-time computable?

\end{document}